\documentclass
[a4paper,10pt,twoside]{cpc-hepnp}

\usepackage{multicol}
\usepackage{graphicx}
\usepackage{booktabs}
\usepackage{amssymb,bm,mathrsfs,bbm,amscd}
\usepackage[tbtags]{amsmath}
\usepackage{lastpage}
\usepackage{CJK}

\usepackage[abs]{overpic}

\begin{document}
\begin{CJK*}{GBK}{song}

\title{A method for the separation and reconstructions of charged hadron and neutral hadron from their overlapped showers in electromagnetic calorimeter
\thanks{Supported by National Natural Science
Foundation of China (No.10435070, No.10773011, No.10721140381,
No.10099630, No.11061140514) and by China Ministry of Science and
Technology(No.2007CB1610, No.2010CB833000)} \footnote{Submitted to
Chinese Physics C} }

\author{%
      LIANG Song$^{1,2}$
\quad TAO Jun-Quan$^{1;1)}$\email{taojq@mail.ihep.ac.cn}%
\quad SHEN Yu-Qiao$^{1,2}$  \quad FAN Jia-Wei$^{1,2}$  \\
\quad XIAO Hong$^{1,2}$ \quad CHEN Guo-Ming$^1$ \quad CHEN
He-Sheng$^1$   \quad BIAN Jian-Guo$^1$ \\ \quad CHEN Ye$^1$ \quad LI
Zu-Hao$^1$  \quad MENG Xiang-Wei$^1$  \quad TANG Zhi-Cheng$^1$  \\
\quad WANG Xian-You$^{1,3}$ \quad WANG Zheng$^1$ \quad XU
Ming$^{1,2}$   \quad XU Wei-Wei$^{1,2}$  \\  \quad YAN Qi$^{1,2}$
\quad YANG Min$^1$
 \quad ZHANG Cheng$^{1,2}$
  \quad ZHU Shi-Hai$^1$
 } \maketitle

\address{%
$^1$ Institute of High Energy Physics, Chinese Academy of Sciences, Beijing 100049, China\\
$^2$ University of Chinese Academy of Sciences, Beijing 100049, China\\
$^3$ Theoretical Physics Institute, Chongqing University, Chongqing 400044, China
}

\begin{abstract}
The separation and reconstructions of charged hadron and neutral
hadron from their overlapped showers in electromagnetic calorimeter
is very important for the reconstructions of some particles with
hadronic decays, for example the tau reconstruction in the searches
for the Standard Model and supersymmetric Higgs bosons at the LHC.
In this paper, a method combining the shower cluster in
electromagnetic calorimeter and the parametric formula for hadron
showers, was developed to separate the overlapped showers between
charged hadron and neutral hadron. Taking the hadronic decay
containing one charged pion and one neutral pion in the final status
of tau for example, satisfied results of the separation of the
overlapped showers, the reconstructions of the energy and positions
of the hadrons were obtained. An improved result for the tau
reconstruction with this decay model can be also achieved after the
application of the proposed method.
\end{abstract}



\begin{multicols}{2}

\section{Introduction}
At LHC, $Higgs\rightarrow\tau\tau$ is an important decay channel in
the searches of Standard Model and supersymmetric Higgs bosons.
About two-thirds of taus decay hadronically to charged hadron(s) and
neutral pion(s), among which about one forth of taus will decay to a
median particle rho($\rho^\pm$) and final one neutrino, one charged
pion and one neutral pion which will immediately decay into two
photons and will reconstructed as a photon candidate in experiment
for the very closure of the two photons. The overlapped shower of 2
photons from a neutral pion will also be thought as a neutral pion
and called as a neutral pion in this paper for convenience. Both the
charged pion and neutral pion will deposit energy in the
electromagnetic calorimeter (ECAL), and their showers will overlap
with each other. So it's very important to know the energy fraction
belong to the neutral pion for the reconstructions of both the
energy and position for the neutral pion in ECAL, which is of course
very important for the reconstructions of $\rho^\pm$ and $\tau^\pm$.
The shower splitting is also important for the reconstruction of
jets which play an important role in the physics analysis of many
hadronic particles giving rise to jets and the measurement of the
missing transverse energy based on jets.

In this paper, a technique was proposed to be used for the
separation and reconstructions of charged hadron and neutral hadron
from their overlapped showers in ECAL, combining the
supercluster\cite{labSC} in ECAL and the empirical formulae for the
parametrization of the hadron showers \cite{labAMSHAD}. The
parametrization of electromagnetic shower and hadronic shower in
electromagnetic calorimeter has been studied in Ref \cite{labAMSEM}
and \cite{labAMSHAD} respectively. The empirical formulae of both
the showers can fit the shower shape well with the data of Alpha
Magnetic Spectrometer II ECAL test beam. And the empirical formula
for the electromagnetic shower was also studied for the
discrimination of unconverted $\gamma/\pi^0$ at the
LHC\cite{labCMSEM}. Taking the channel
$\tau^{\pm}\to\rho^{\pm}+\nu_{\tau^{\pm}}\to\pi^{\pm}+\pi^0+\nu_{\tau^{\pm}}$
for example in this paper, the empirical formula for the hadronic
shower was used in the process of the separation and reconstructions
of charged hadron and neutral hadron from their overlapped showers.
The results shown that the overlapped can be well split and the
energy can be well assigned to the charged pion and neutral pion.
And the median particle can be well reconstructed with an improved
result than the present reconstruction algorithm used in CMS
experiment at LHC\cite{labPF}.

The setup of the detector with the GEANT4 package and the
Supercluster algorithm in electromagnetic calorimeter are simply
introduced in section 2. The simple description and validation of
the parametric shower shape formula for hadron shower are described
in section 3. The proposed technique and its performance are
presented in section 4. Finally the summary and outlook are in
Section 5.

\section{ECAL and SuperCluster algorithm description}

\subsection{Electromagnetic Calorimeter}

In particle experiment, electromagnetic calorimeter is one
specifically designed to measure the energy of photons and
electrons, which will deposit their energy in ECAL after the
electromagnetic interaction and showering in ECAL. As Ref
\cite{labCMSEM}, we constructed an ECAL geometry with GEANT4
package, as the ECAL barrel region of CMS detector, to study the
shower of electromagnetic style particles and develop the separation
and reconstruction algorithm of hadrons from the overlapped shower
in ECAL with the empirical formula for charged hadron shower.

The ECAL barrel is made of 61200 same-sized lead tungstate(PbWO4)
crystals, covering pseudorapidity $|\eta|<1.479$. The centers of the
front faces of the crystals are at a radius 1.29m from the global
coordinate $z-axis$. The cross-section of the front face and rear
face for each crystal is corresponds to $0.0174\times0.0174$ (in
unit of radian$\times$radian) in $\eta-\phi$ plane, and each crystal
230mm long which is corresponding to 25.8X0 with 1 radiation
length(X0) equal to 0.89 cm. They are mounted in a quasi-projective
geometry so that their axes make a small angle ($3^o$) with respect
to the vector from the nominal interaction point (i.e, the center of
the detector setup), in both the $\eta$ and $\phi$ directions. So in
$\eta-\phi$ plane, the barrel crystals project approximatively to be
a series of $360\times170$ squares, with which the lateral shower of
a single charged hadron can be described well by the empirical
formula as described in section 2. Additional the same tracker
detector located in the front of ECAL and magnetic field with 3.8
Tesla along the $\eta$ direction are also constructed to measure the
momentum of the track from charged hadron. More detail descriptions
can be found in Ref\cite{labCMSDet}.

\subsection{Supercluster algorithm}

Photons and electrons will shower in ECAL and deposit about 94\% of
their energy in $3\times3$ crystals, and $97\%$ in $5\times5$
crystals. Especially for unconverted photons and electrons with less
bremsstrahlung in test beam, the fixed arrays can give a much better
performance. But the presence of tacker material in front of ECAL,
results in bremsstrahlung of electron and photon conversions. So the
energy deposited from electrons and photons showering in the
calorimeter will spread in $\phi$ direction due to the strong
magnetic field. The hybrid supercluster algorithm \cite{labSC} is
developed to collect the separated and the bremsstrahlung energy in
the $\phi$ direction. The algorithm start from the find of a seed
crystal with deposit energy greater than 0.35GeV. The $1\times3$
($\eta\times\phi$) or if $1\times3$ energy greater than 1GeV then
$1\times5$ crystal arrays in $\eta$ direction will be included to
the same supercluster, until the energy in $1\times3$ less than
0.1GeV or at most including 11 such arrays. This kind of
supercluster, in a very narrow $\eta$ window and much wider $\phi$
window, gives very nice performance for electrons and photons in the
ECAL as we described in the above section. Two photons from a
neutral pion decay can deposit their energy very close  to each
other to result in the overlapped shower can be reconstructed as one
hybrid supercluster. The position of the center of gravity (COG) of
the supercluster in ECAL can be calculated with a good position
resolution using the energy and position of each crystal contained
in this supercluster, as decribed in Ref \cite{labSC} and Ref
\cite{labCMSEM}. More details about the algorithm can be found in
Ref \cite{labSC}.

\section{Validation of the parametric formula of hadronic shower}

\subsection{Description of the parametric shower shape formula}

The longitudinal shower formula and lateral formula of hadronic
shower were studied in Ref\cite{labAMSHAD}. In the homogenous
crystal as we constructed in GEANT4, the shower shape of single
charged hadron particle is symmetrical around the shower direction.
So the lateral shower in the transverse plane vertical to the
particle direction is isotropic. In this paper, only the lateral
shower formula was used to describe the shower shape due to the
difficulty to find out the shower start point of the hadronic shower
in the constructed ECAL with only one crystal from the front to the
rear. We can estimate the position of the center of gravity (COG) of
supercluster in ECAL of the cascade shower using the energy and
position of each crystal with a satisfied position
resolution\cite{labSC}\cite{labCMSEM}. Through the COG point we can
make a transverse plane which is vertical to the shower direction.
Then the whole shower shape will be projected to this plane. The
deposited energy density can be described by the following
parameterized function,
%which is the transformed formula in the
%system of polar coordinates as the formula developed in
%Ref\cite{labAMSHAD} with parameter $B=3$:

\begin{equation}
\label{eq1} \frac{dE}{dr}=f(E,R,r)=2Er\frac{R}{(r+R)^3},
r=\sqrt{(x-x_c)^2+(y-y_c)^2},
%\label{eq1}
%\frac{d^2E}{dxdy}=f(E,R,r)=\frac{E}{\pi} \frac{R}{(r+R)^3},
%r=\sqrt{(x-x_c)^2+(y-y_c)^2},
\end{equation}
which is the transformed formula in the system of polar coordinates
as the following formula developed in Ref\cite{labAMSHAD} with
parameter $B=3$,
\begin{equation}
\label{eqAMS}
\frac{d^2E}{dxdy}=\frac{E}{2\pi}\frac{\Gamma(B)}{\Gamma(B-2)}\frac{R^{B-2}}{(r+R)^B}.
%r=\sqrt{(x-x_c)^2+(y-y_c)^2}, \label{eq1}
%\frac{d^2E}{dxdy}=f(E,R,r)=\frac{E}{\pi} \frac{R}{(r+R)^3},
%r=\sqrt{(x-x_c)^2+(y-y_c)^2},
\end{equation}
%where
In these formulae, $r$ is the distance of the shower developing
point to COG in transverse plane. $E$ is the total deposited energy
of the shower. $R$ is a free parameter to describe the shower shape.
We will validate the parametric formula with charged pion showers in
the constructed ECAL in the next subsection.

\subsection{Verification and parameter determination}

Single charged pion runs with incident energy ranging from 20 to 100
GeV were used to verify the formula. When the total deposited energy
in ECAL is larger than 1 GeV, the energy of $5\times5$ crystals
around the maximum energy crystal as Fig.~\ref{fig1}(a), were fitted
with formula \ref{eq1} using $\chi^2$ minimization method in MINUIT
package. The fit minimized $\chi^2$ is given by
\begin{equation}
\label{eq2}
 \chi^2 = \sum^{25}_{i=1}(E^{i}_{fitted} - E^{i}_{deposited})^2
\end{equation}
where $E^{i}_{deposited}$ is the energy deposited in $ith$ cell;
$E^{i}_{fitted}$ is the energy in $ith$ cell predicted by the
formula \ref{eq1}. The integration algorithm for the formula along
$r$ direction is the same as described in Ref \cite{labCMSEM}.

From the study, more than 98\% of the charged pion showers in ECAL
were fitted  successfully with formula \ref{eq1}. The fitting
results including the shower shape in $5\times5$ cell array,
$\chi^2/E^2$ and $R$ distributions are shown in Fig.~\ref{fig1}.
From the plot Fig.~\ref{fig1}(b), the agreement between the shape
calculated by formula \ref{eq1} and the original deposited energy
for the hadronic shower is satisfied. And the maximum energy cell
contain about 52\% of the total energy. From Fig.~\ref{fig1}(d), the
value of the parameter $R$ is about 0.25 with a maximum probability.

\end{multicols}
\ruleup
\begin{center}
\begin{minipage}[t]{0.49\textwidth}
\centering
%\begin{overpic}[scale=0.36]{CrystalNumber.eps}
\begin{overpic}[width=0.49\textwidth,height=0.49\textwidth]{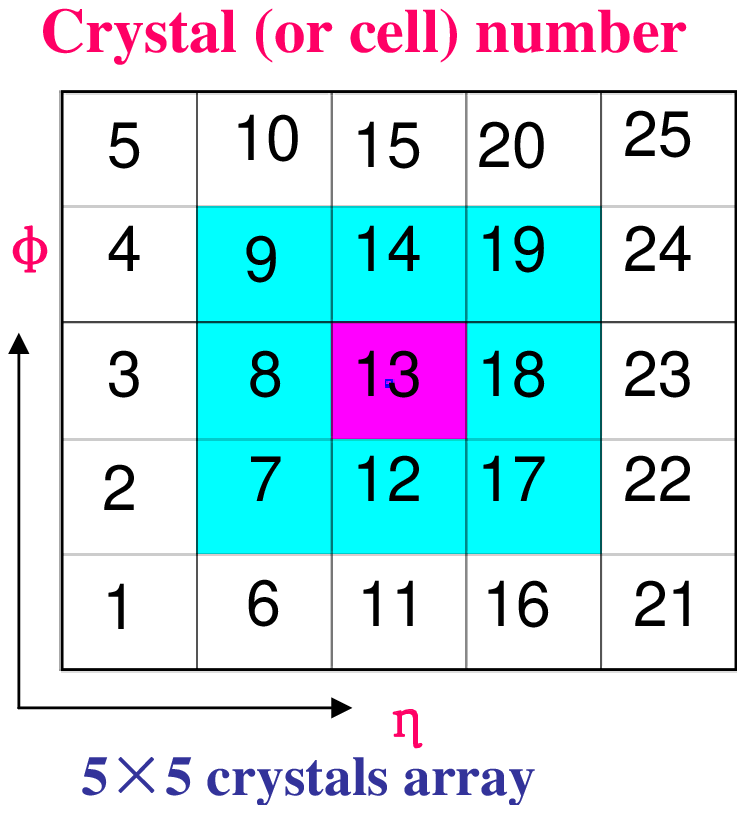}
\put(90,0){$(a)$}
\end{overpic}
\hfill
\end{minipage}
\begin{minipage}[t]{0.49\textwidth}
\centering
\begin{overpic}[scale=0.36]{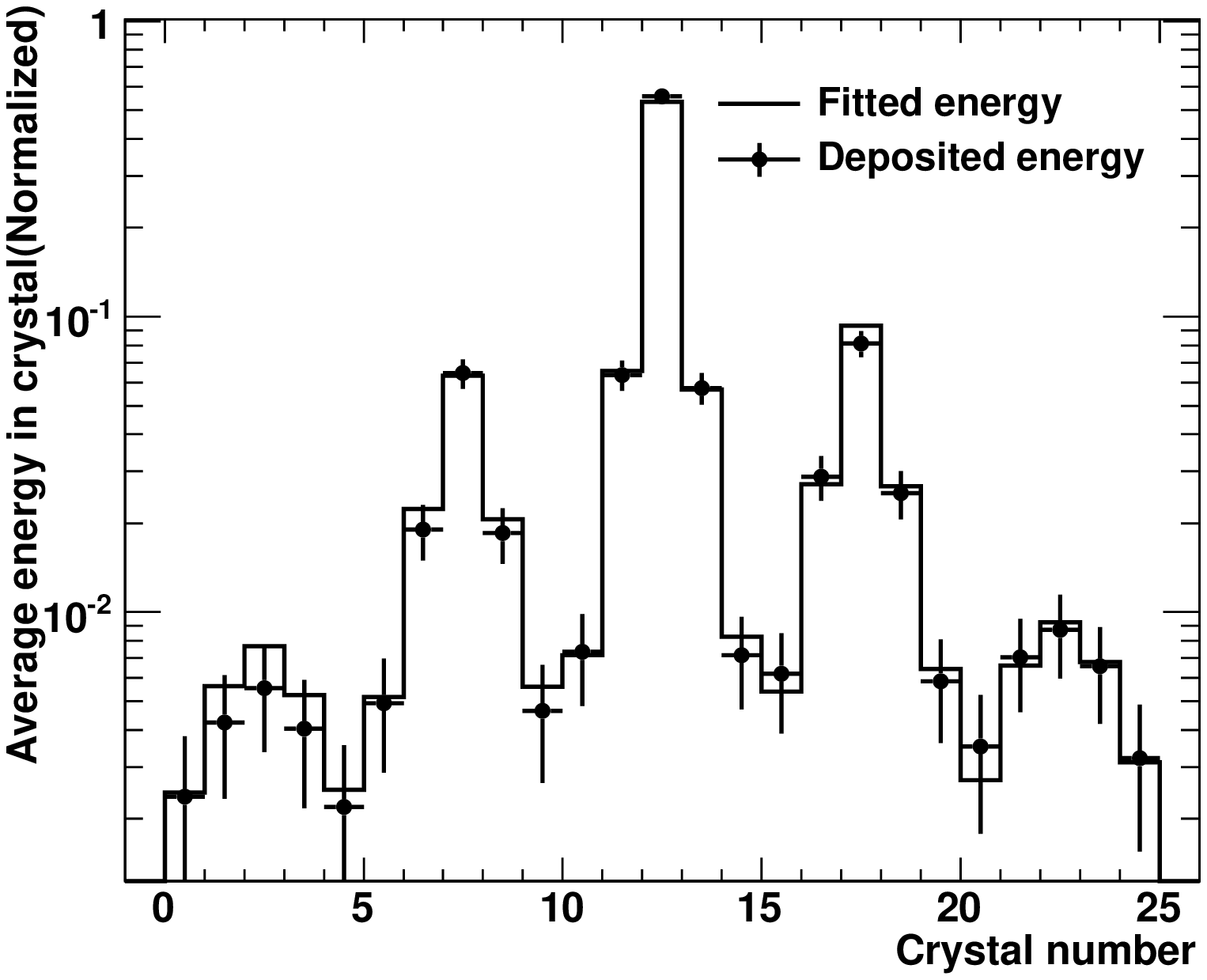}
\put(90,0){$(b)$}
\end{overpic}
\end{minipage}
\begin{minipage}[t]{0.49\textwidth}
\centering
\begin{overpic}[scale=0.36]{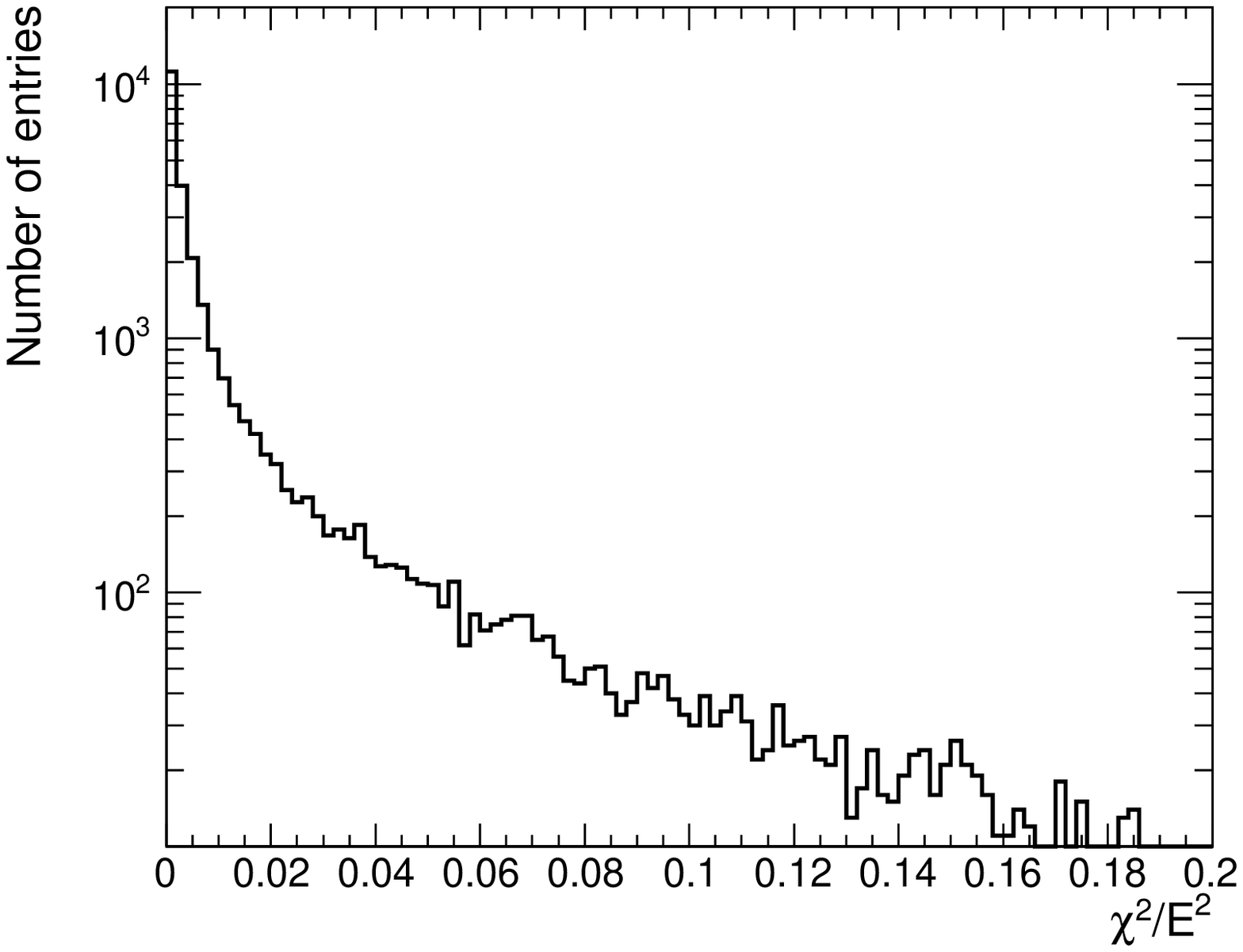}
\put(90,0){$(c)$}
\end{overpic}
\end{minipage}
\begin{minipage}[t]{0.49\textwidth}
\centering
\begin{overpic}[scale=0.36]{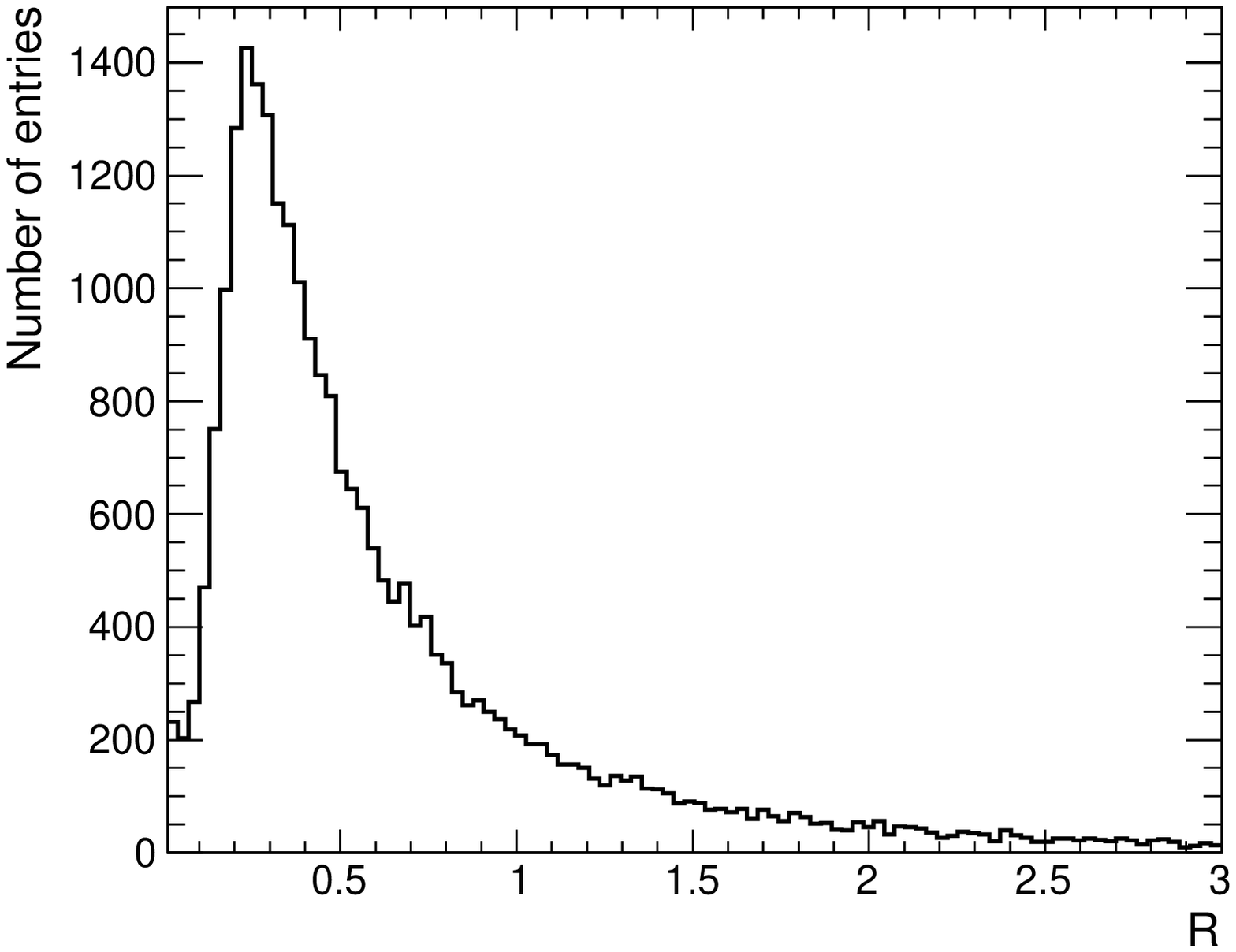}
\put(90,0){$(d)$}
\end{overpic}
\end{minipage}
\figcaption{ (a) The numbering of $5\times5$ crystals array. $\ $
(b) Comparisons of the average predicted energy with the average
original deposit energy in each crystal of the $5\times5$ crystals
array for 20-100GeV charged pion samples. The dots represent the
average original deposit energy in each crystal and the histogram
represents the predicted one by formula\ref{eq1}.$\ $ (c) The
$\chi^2/E^2$ distribution represent the goodness of the fitting.$\ $
(d) The distribution of the only parameter $R$ in formula\ref{eq1}.}
{\label{fig1}}
\end{center}
\ruledown

\begin{multicols}{2}

\section{Method for separation and reconstruction of neutral pion and charged pion from their overlapped shower in ECAL}
\subsection{Description of the method}
As described at the very beginning in the first section, for many
physics objets in experiment such as the reconstructions of tau and
jet, the final neutral pions and charged pions will produced very
closely with each other in higher energy hadron collider. Their
showers in ECAL will overlapped with each other and will be
reconstructed as one supercluster in experiment. We tried to use the
shower shape method with the parametric formula as described above
to separate the overlapped shower and then obtain the energy and
position of charged pion and neutral pion. This method is also taken
as one of the applications of the hadronic shower shape formula in
Ref \cite{labAMSHAD}.

We take the most simple decay module of tau,
$\tau^{\pm}\to\rho^{\pm}+\nu_{\tau^{\pm}}\to\pi^{\pm}+\pi^0+\nu_{\tau^{\pm}}$
which is almost one forth of the tau decays, to describe the method.
Fig.~\ref{fig2} is the schematic view of the shower overlapping of
the neutral hadron and charged hadron. The two photons from a
neutral pion are close to each other and will form the EM shower.
The hadronic shower in ECAL from $\pi^\pm$ shower will also be very
close to the EM shower. They will be reconstructed as one
superclusters in ECAL. With the formula \ref{eq1}, we can calculate
the hadron energy in each crystal, if we have the values of the
following three variables, COG(center of gravity), total energy E,
and parameter R.  So for the method, firstly the fix or the
estimation of their values will be considered in the following .

\begin{center}
\includegraphics[width=0.45\textwidth]{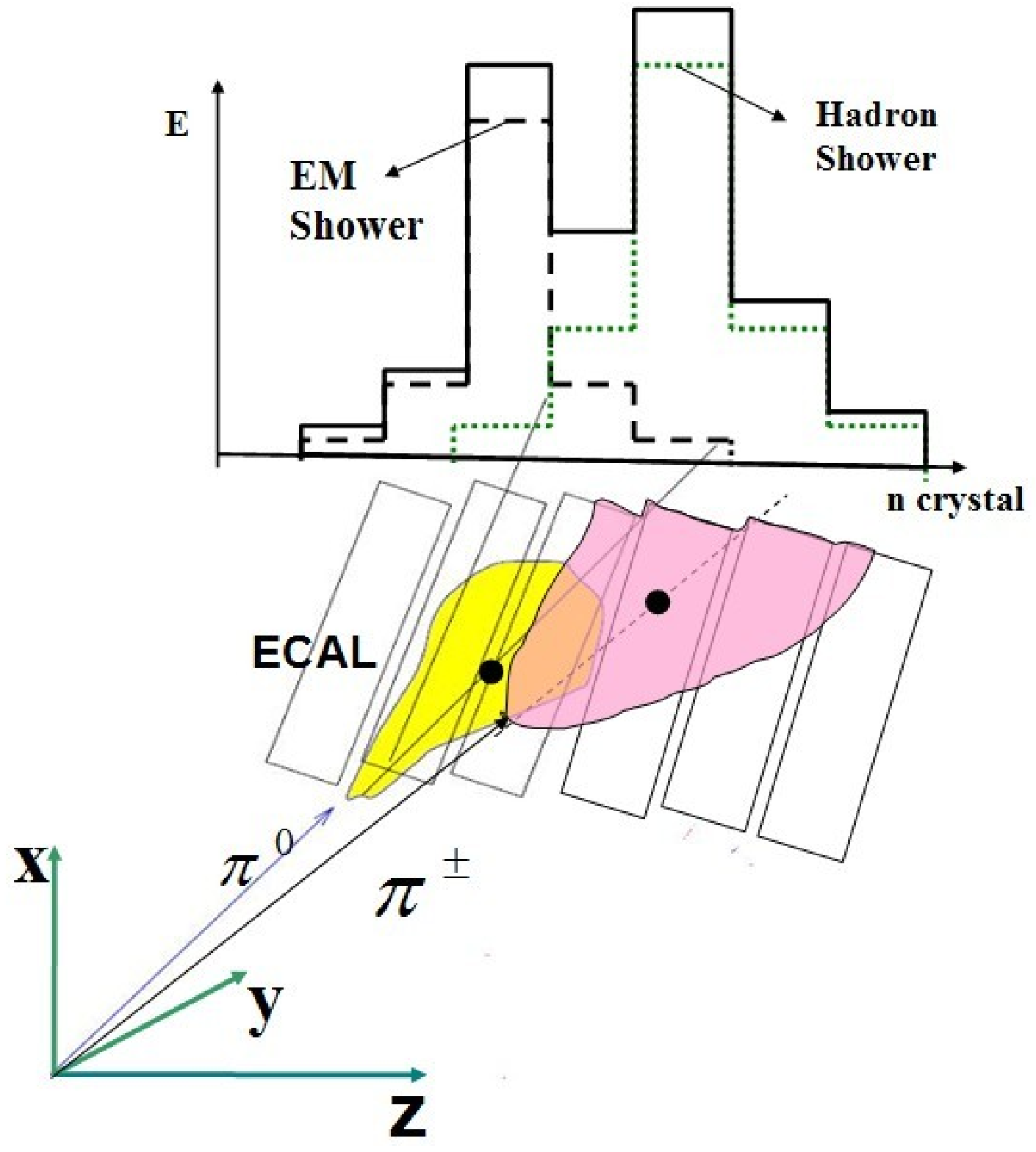}
\figcaption{\label{fig2}   Schematic view of the shower method}
\end{center}

From Fig.~\ref{fig1} (d), $R=0.25$ is applied in the formula for it
has a peak value at about 0.25. The COG of $\pi^\pm$ shower should
be close to the track reconstructed in the tracker material and
extrapolated to the front face of ECAL. The radius of COG is around
137.0cm to the the center of the constructed detector in section 2,
as shown in Fig.~\ref{fig3}(a). So the point on the track at
$radius=137.0 cm$ is chosen to be taken as the COG of the hadronic
shower. For the shower from single particle, the COG and be also
calculated from the algorithm described in section 2.2 or in Ref.
\cite{labCMSEM}. The difference of $\eta$ and $\phi$ between these
two points are shown in Fig.~\ref{fig3}(b)(c). The $\sigma$ of the
difference between the reconstructed one and the truth is much lower
than one crystal size as described in section 2. The maximum hit
energy of $\pi^\pm$ shower is used to estimate the total deposit
energy $E$, for it contains about 52\% of $\pi^\pm$ energy as shown
in Fig.~\ref{fig1}(b). Now the key is to find the crystal which has
maximum energy of hadron shower. The crystal with the largest
deposited energy in the all crystals with which the track will path
through in ECAL, not always the most energic crystal in the shower
was chosen as the maximum energy of hadron shower. The differences
of crystal number between selected crystal for the seed of hadron
and the most energic crystal in the shower is given in
Fig.~\ref{fig3}(d) and (e) in the $\eta$ and $\phi$ direction
respectively. From the plots, the probability of the selected
crystal for the seed of hadron being also the most energic crystal
in the shower is about 91\% of all $\pi^\pm$ events both in the
$\eta$ and $\phi$ directions.
%The difference between the estimated total energy and
%the truth one is shown in Fig.~\ref{fig3}(f).

\end{multicols}
\ruleup
\begin{center}
\begin{minipage}[t]{0.32\textwidth}
\centering
\begin{overpic}[scale=0.30]{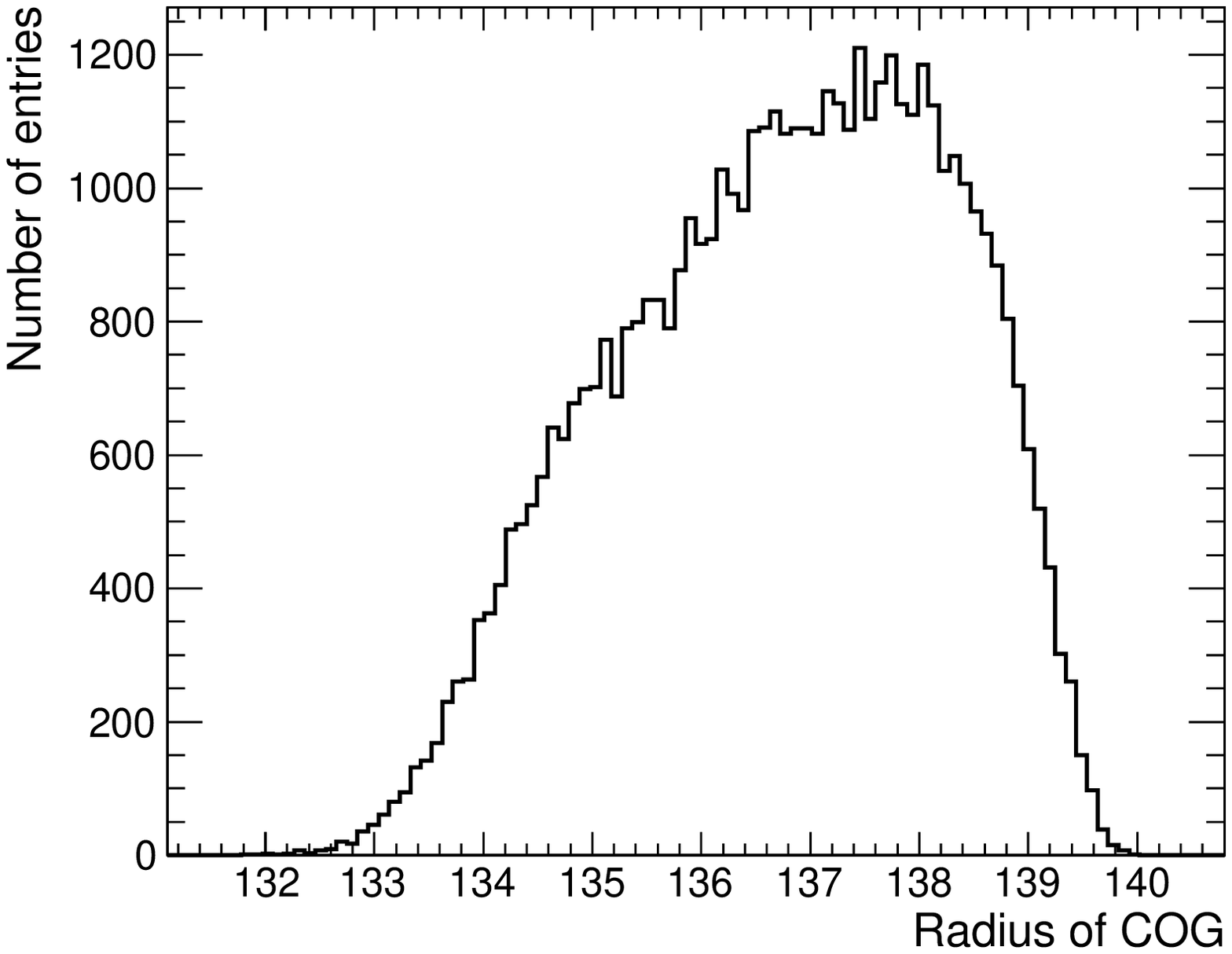}
\put(90,0){$(a)$}
\end{overpic}
\end{minipage}
\begin{minipage}[t]{0.32\textwidth}
\centering
\begin{overpic}[scale=0.30]{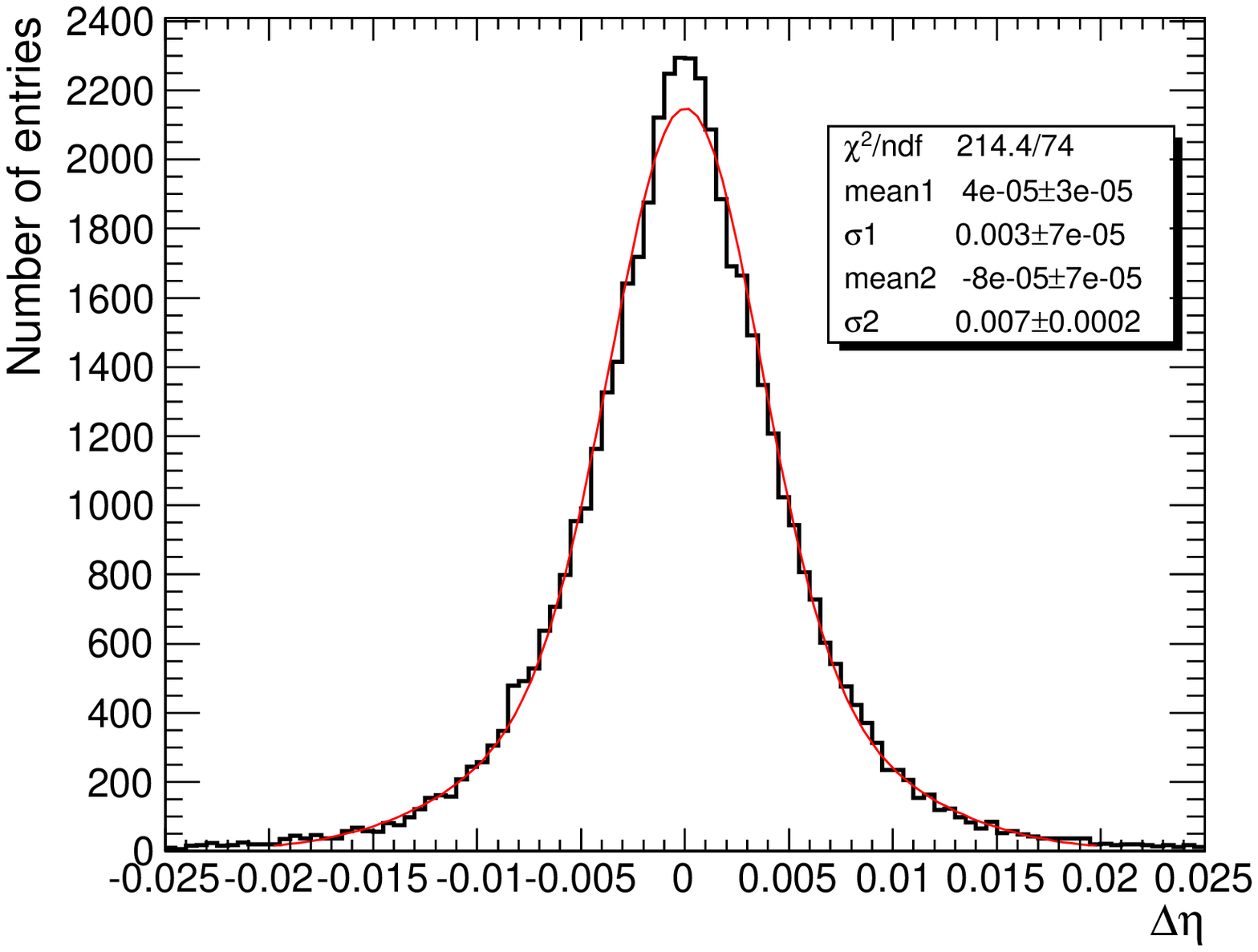}
\put(90,0){$(b)$}
\end{overpic}
\end{minipage}
\begin{minipage}[t]{0.32\textwidth}
\centering
\begin{overpic}[scale=0.30]{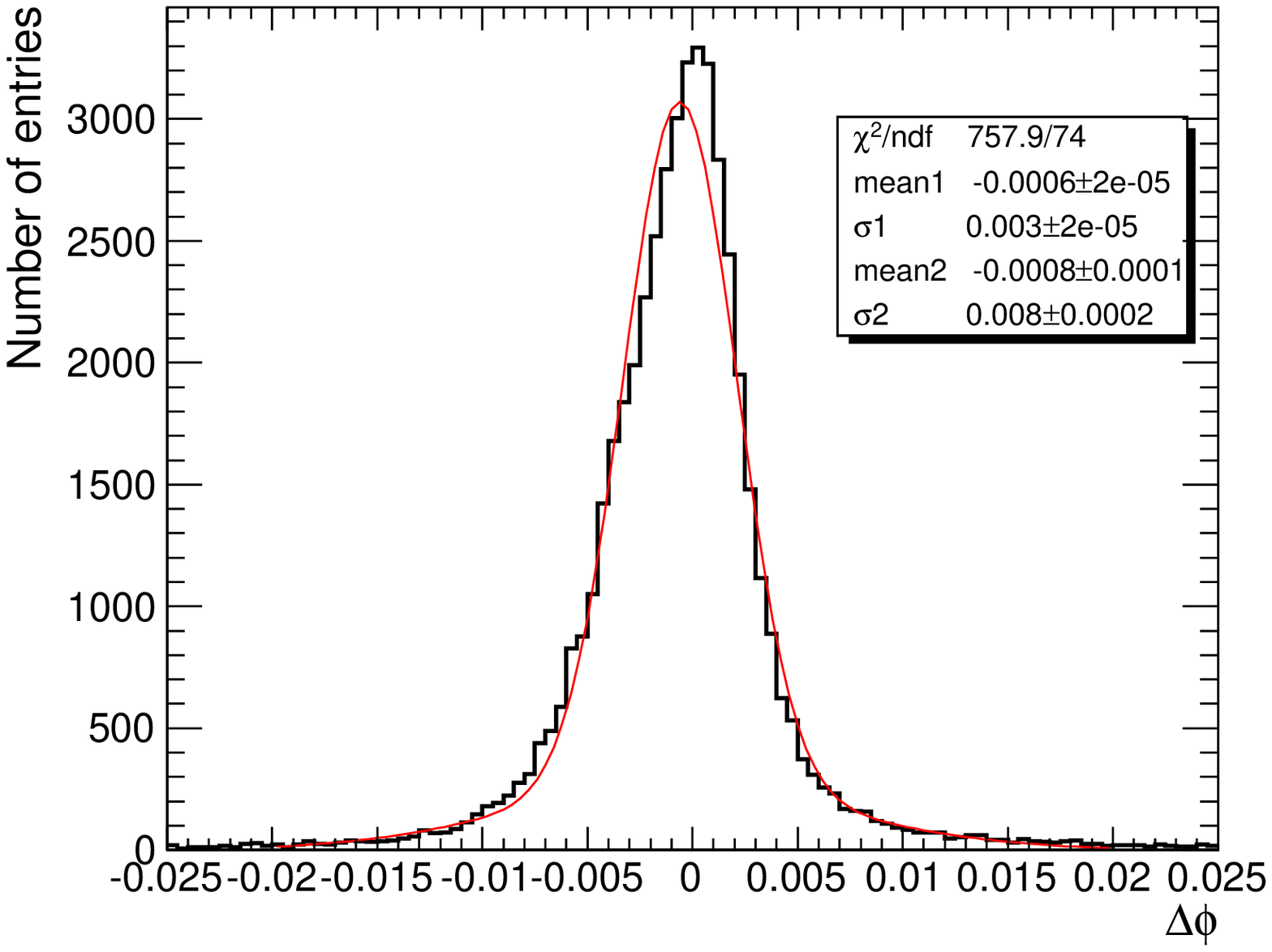}
\put(90,0){$(c)$}
\end{overpic}
\end{minipage}
\begin{minipage}[t]{0.32\textwidth}
\centering
\begin{overpic}[scale=0.30]{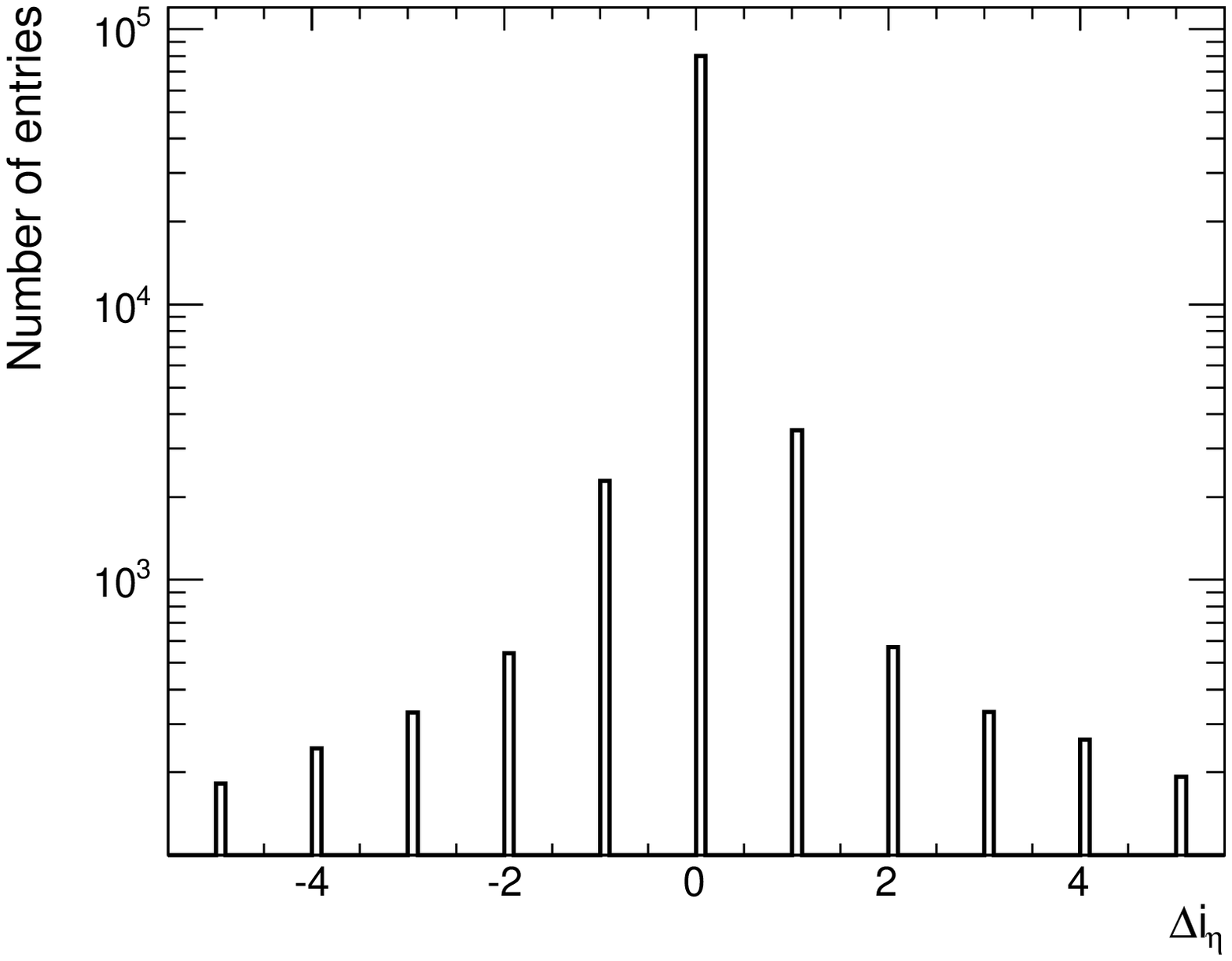}
\put(90,0){$(d)$}
\end{overpic}
\end{minipage}
\begin{minipage}[t]{0.32\textwidth}
\centering
\begin{overpic}[scale=0.30]{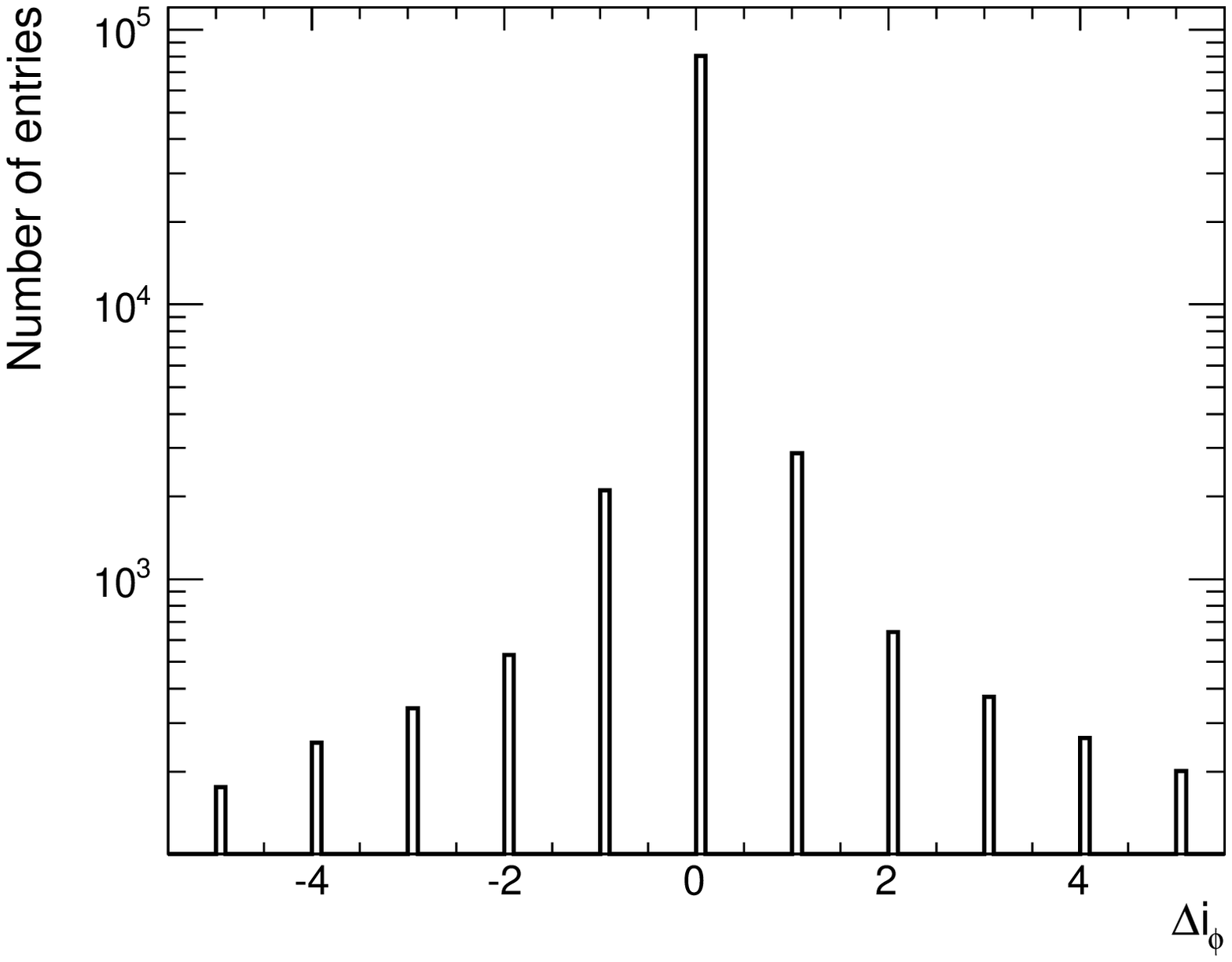}
\put(90,0){$(e)$}
\end{overpic}
\end{minipage}
%\begin{minipage}[t]{0.32\textwidth}
%\centering
%\begin{overpic}[scale=0.30]{dE.eps}
%\put(90,0){$(f)$}
%\end{overpic}
%\end{minipage}
\figcaption{(a) Radius distribution of the COG of the charged hadron
showers from $\pi^\pm$. (b) Differences of $\eta$  and (c) $\phi$
between COG and its estimate point on the track of the incident
$\pi^\pm$. (d) Differences of crystal number between maximum hit
crystal and its estimate in $\eta$ direction and (e) in $\phi$
direction for the single charged hadron shower from $\pi^\pm$.
% and (f) difference of $\frac{\Delta E}{E}$ between
%the estimated energy and its truth one from the incident $\pi^\pm$.
}{\label{fig3}}
\end{center}
\ruledown

\begin{multicols}{2}

Now the process of the method is introduced here. There are several steps for the separation and reconstruction
of neutral pion and charged pion from their overlapped shower, the hybrid supercluster in ECAL: \\
1) Firstly we extrapolated the $\pi^\pm$ track to the front face of ECAL. The
energy ($E_{maxHit}$) of the crystal which has the largest deposited energy the tack passed through in ECAL,
was selected as the center of the hadronic shower from the charged hadron. Then the total deposited energy
from the charged pion in ECAL was estimated as $E=E_{maxHit}/0.52$. \\
2) The position on the track at $radius=137.0cm$ was taken as the center of gravity of the shower from charged pion. \\
3) The energy deposited in each crystal from the charged pion can be calculated by the formula \ref{eq1} for the hadronic shower as describe in the above section. \\
4) The remain energy in each crystal after the substraction of the
energy deposited by charged pion  and calculated in 3),
was taken as the deposited energy of the EM shower from the neutral pion. \\
5) The energy of $\pi^0$ can be obtained from the remain energy in
each crystal of the overlapped supercluster. And the position of
$\pi^0$ can be also calculated from the remain energy of the
position of each crystal using the energy weighted algorithm as
described in the Ref \cite{labCMSEM}.

\subsection{Results of the method}
Combining the supercluster and the lateral hadron shower formula
with several approximate calculations, the deposited energy and
position of $\pi^\pm$ and $\pi^0$ from the overlapped shower in ECAL
can be well reconstructed. With the method described above, the
energy and position of charged pion and neutral pion were compared
with the truth energy and position of $\pi^\pm$ and $\pi^0$
respectively, as shown in Fig.~\ref{fig4}(a) (b) and (c). Since all
the overlapped showers were contained during the analysis, including
the energy loss in ECAL by charged pion as a minimum ionizing
particle(MIP), the resolutions of the position and energy of $\pi^0$
were some little bit of large from our method. But the result is
satisfied for us to solve the shower overlapping problem in ECAL for
the first time to use such a technique. This will result a much
better reconstruction of the higher object such as $\rho^\pm$ and
jets. The reconstructed position and energy of the charged pions
from the split shower method were also be compared with the true
ones in Fig.~\ref{fig4}(d) (e) and (f), although the momentum and
position from the tracks were used for the further analysis in the
next paragraph. The results of the position and energy of charged
hadron from the estimated energy in ECAL after the shower splitting
are also satisfied.

\end{multicols}
\ruleup
\begin{center}
\begin{minipage}[t]{0.32\textwidth}
\centering
\begin{overpic}[scale=0.30]{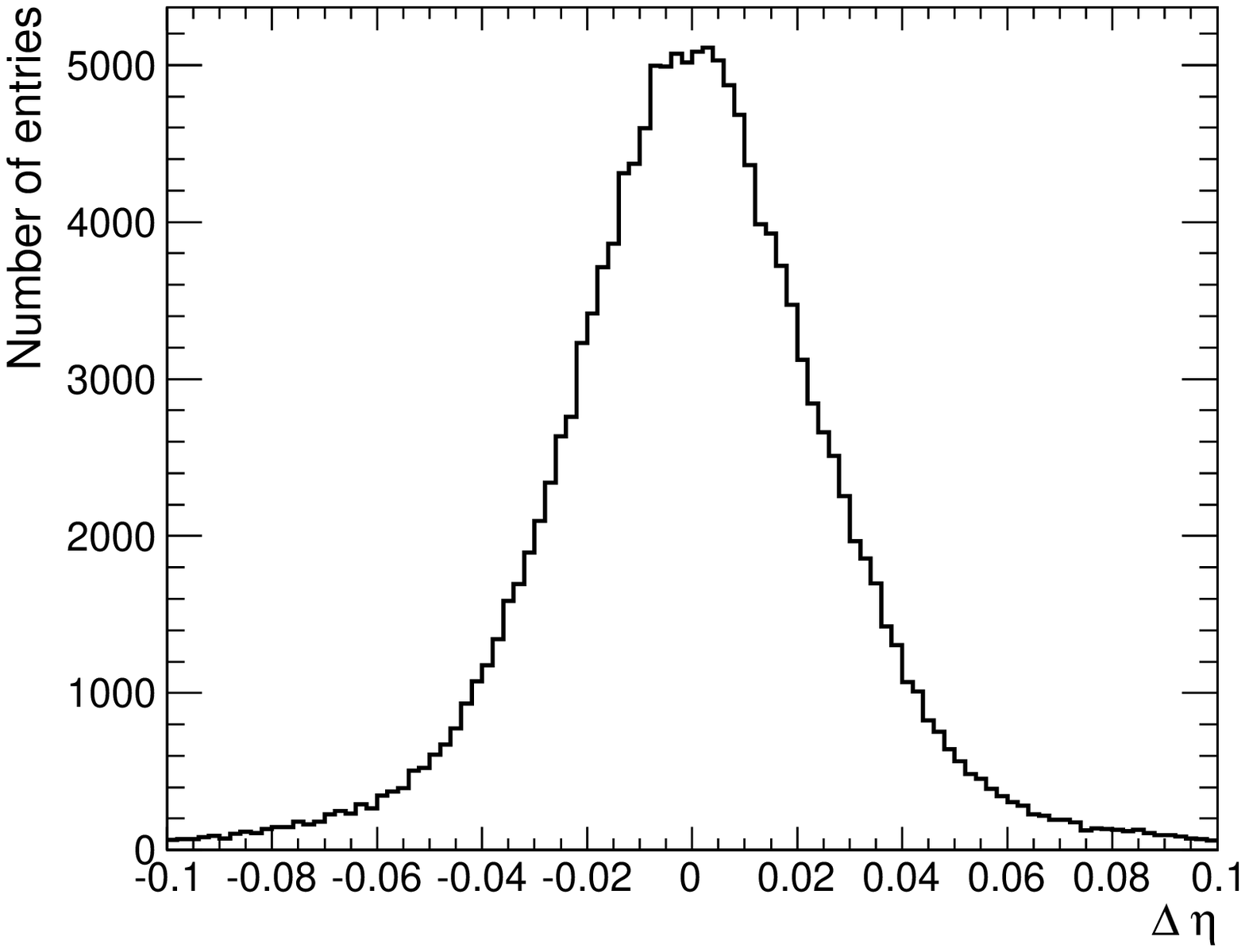}
\put(90,0){$(a)$}
\end{overpic}
\end{minipage}
\begin{minipage}[t]{0.32\textwidth}
\centering
\begin{overpic}[scale=0.30]{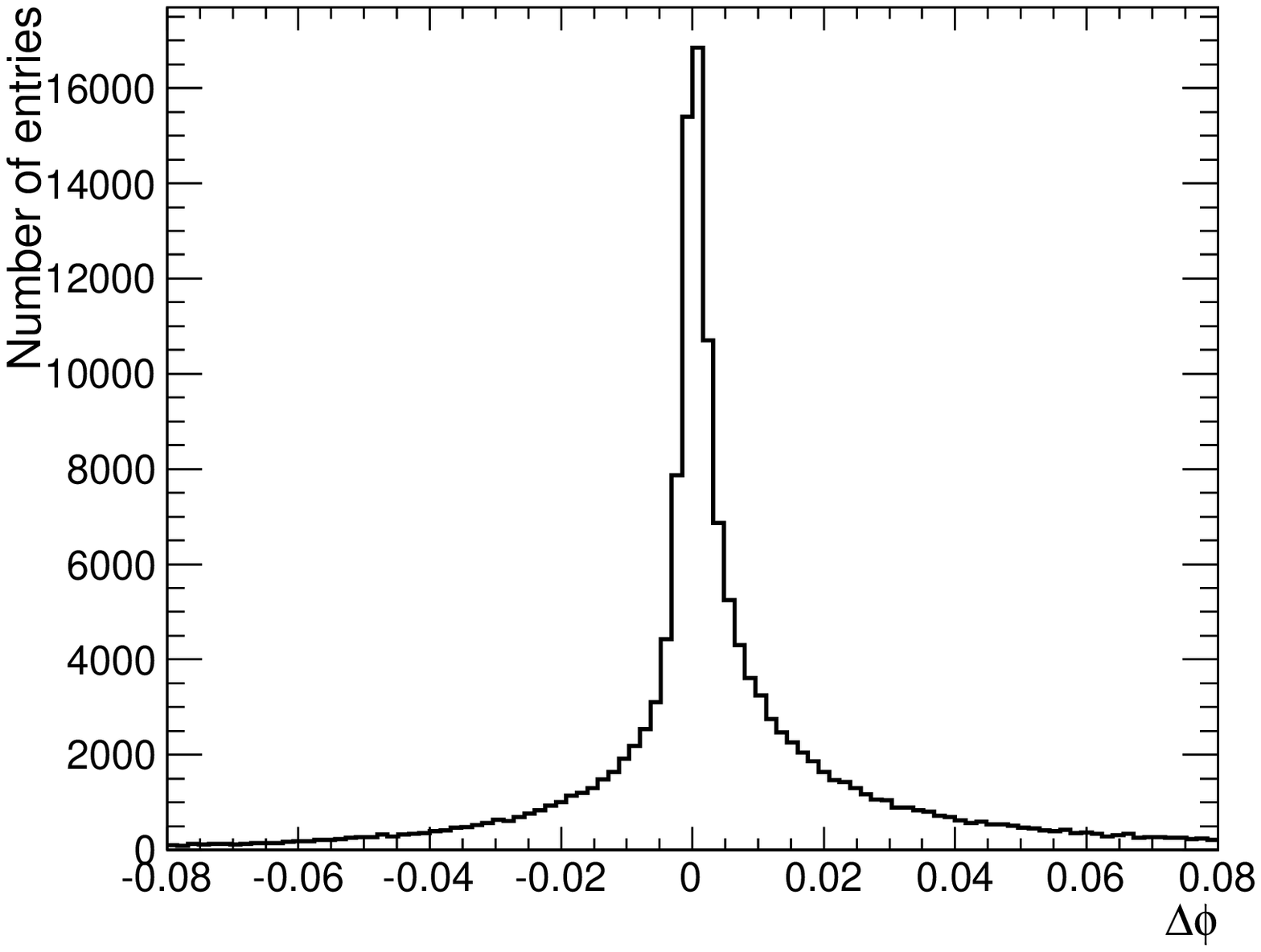}
\put(90,0){$(b)$}
\end{overpic}
\end{minipage}
\begin{minipage}[t]{0.32\textwidth}
\centering
\begin{overpic}[scale=0.30]{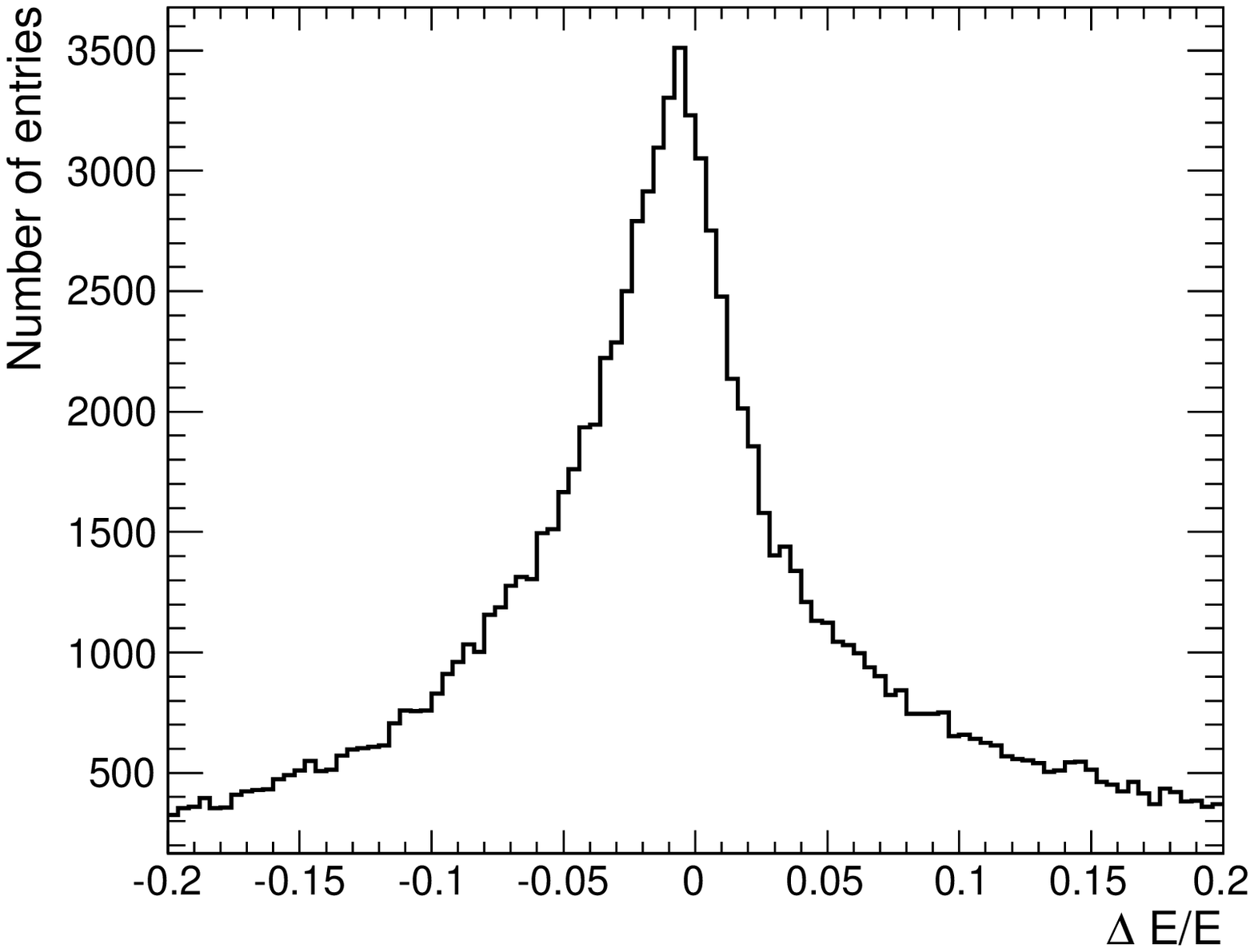}
\put(90,0){$(c)$}
\end{overpic}
\end{minipage}
\begin{minipage}[t]{0.32\textwidth}
\centering
\begin{overpic}[scale=0.30]{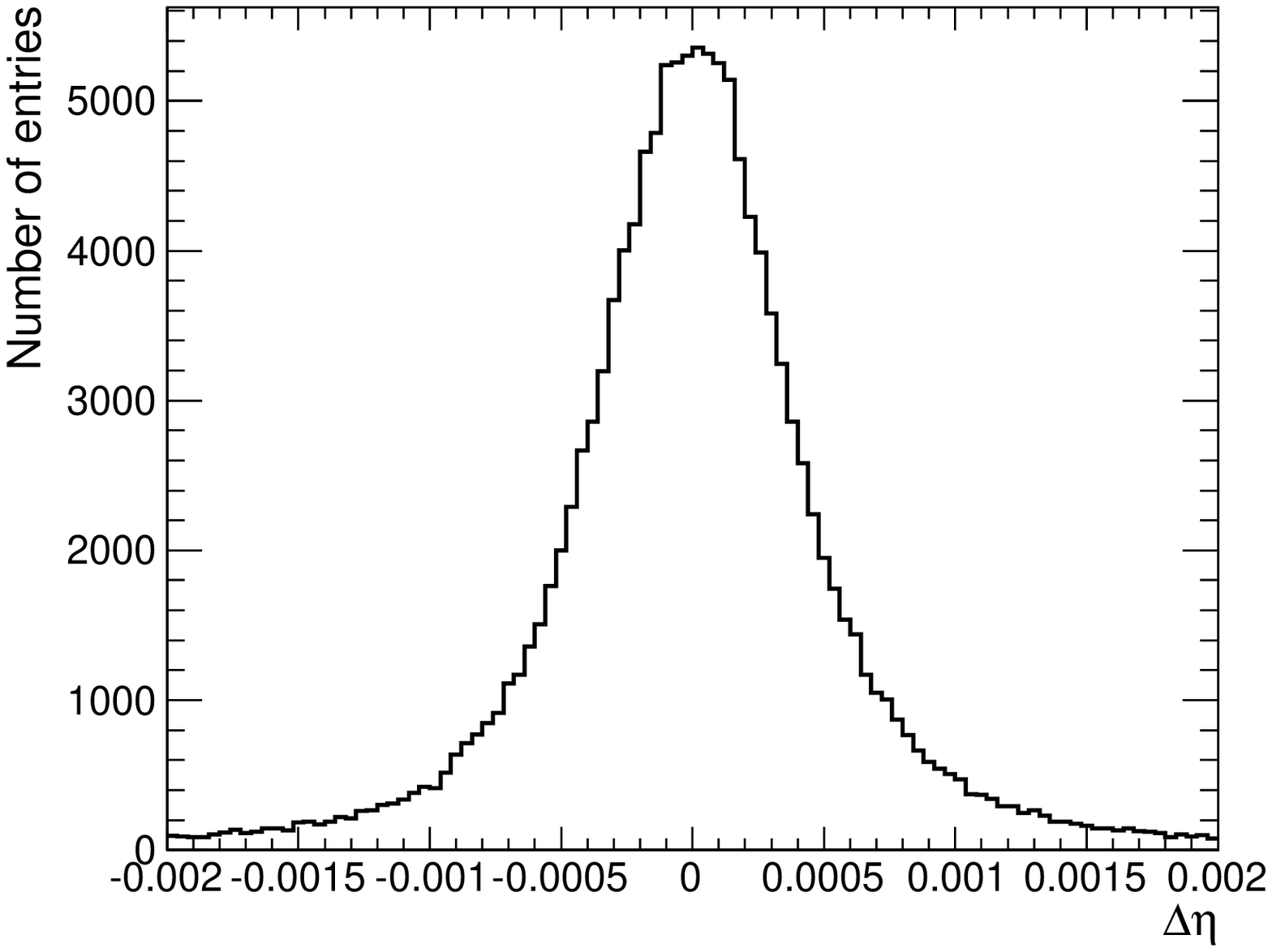}
\put(90,0){$(d)$}
\end{overpic}
\end{minipage}
\begin{minipage}[t]{0.32\textwidth}
\centering
\begin{overpic}[scale=0.30]{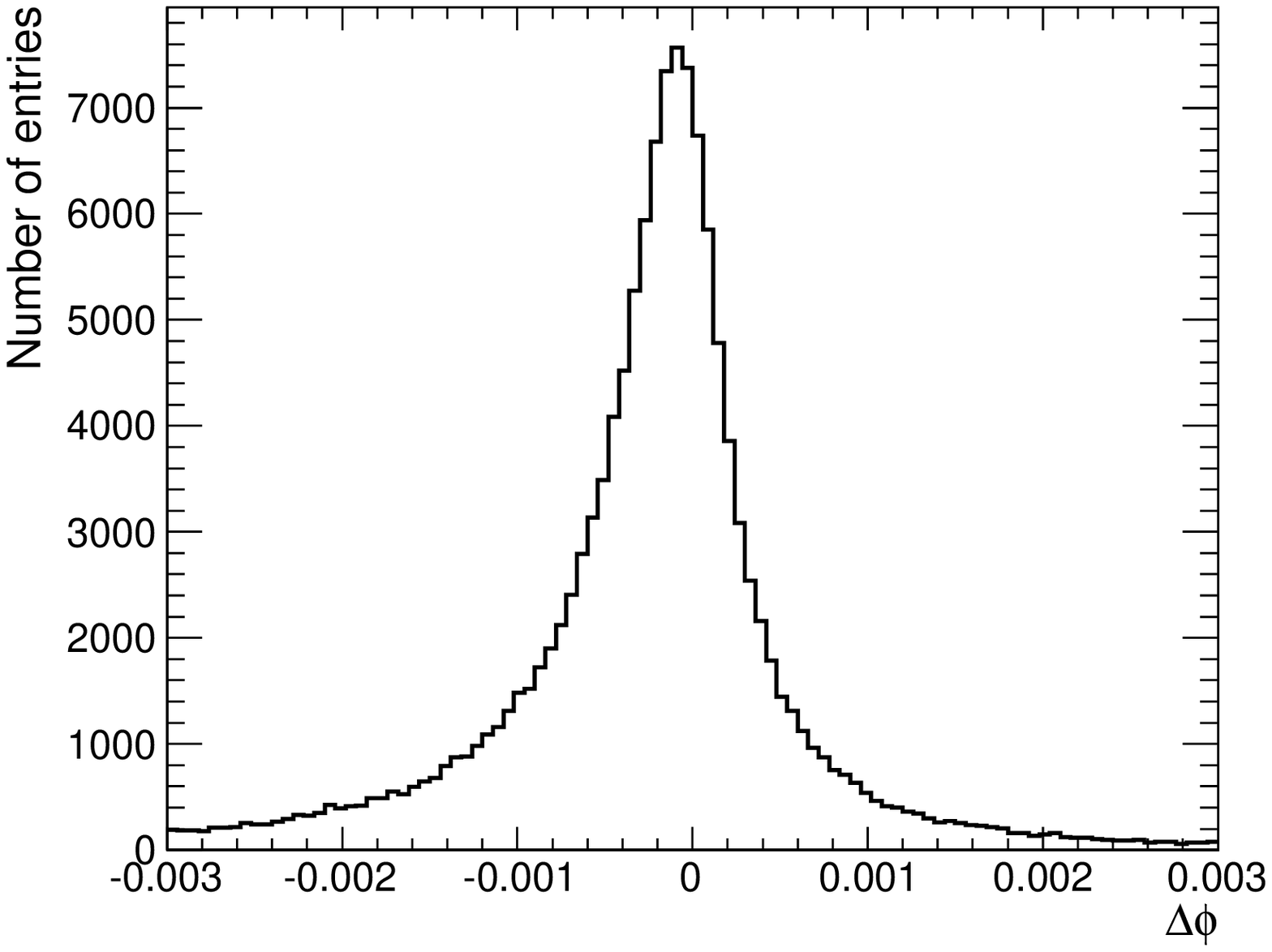}
\put(90,0){$(e)$}
\end{overpic}
\end{minipage}
\begin{minipage}[t]{0.32\textwidth}
\centering
\begin{overpic}[scale=0.30]{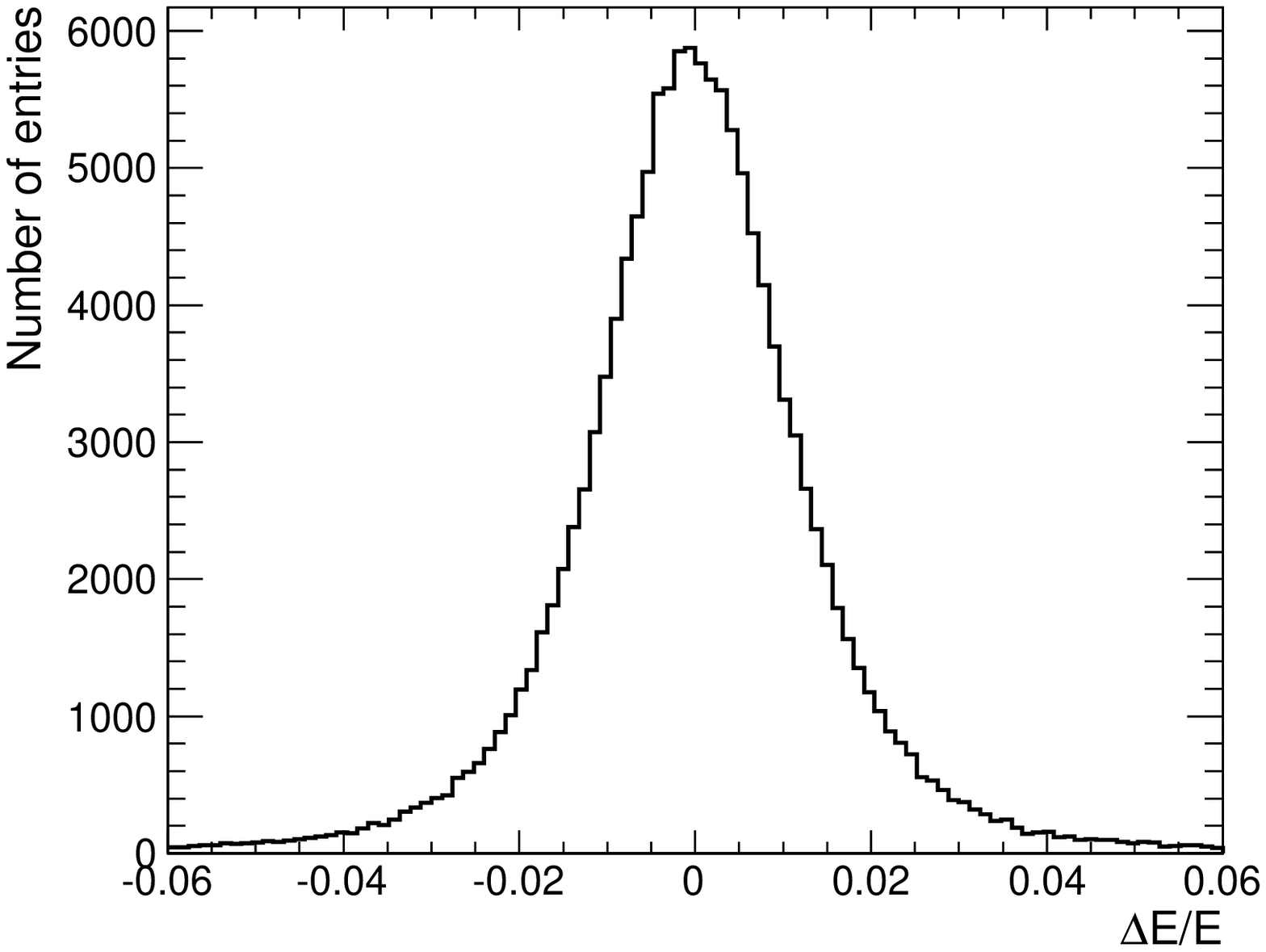}
\put(90,0){$(f)$}
\end{overpic}
\end{minipage}
\figcaption{The performance of $\pi^0$ and $\pi^\pm$ reconstruction
from the split showers after the application of the proposed method
in this pare. (a) The differences of position in $\eta$ and (b)
$\phi$, and (c) energy $\frac{\Delta E}{E}$ between the
reconstructed $\pi^0$ with the separation method and the truth one.
(d) The differences of position in $\eta$ and (e) $\phi$, and (f)
energy $\frac{\Delta E}{E}$ between the reconstructed $\pi^\pm$ with
the separation method and the truth one.}{\label{fig4}}
\end{center}
\ruledown

\begin{multicols}{2}

The reconstructed position and mass of median particle $\rho^\pm$,
i.e., the visible position and mass of $\tau^\pm$ in experiment can
be obtained. During the calculation, the momentum and position from
the tack were used for the charged pion. As comparisons, the results
from the HPS(hadron plus strip) algorithm \cite{labPF}. which is
being used in the present reconstruction software in CMS experiment
at the LHC, were also dawn in the same plot,  as seen from
Fig.~\ref{fig5}. The HPS algorithm include firstly the recalibration
of the energy of ECAL shower using MIP events and then the
comparison between the momentum of tracks and the corresponding
energy in electromagnet plus the hadronic calorimeters to separate
the energy of $\pi^\pm$ and $\pi^0$. From the comparisons, we can
see that improved results can be obtained after the application of
the method for separation and reconstruction of neutral pion and
charged pion from their overlapped shower in ECAL. We can get a
little better results for the position and energy reconstructions
and much better result of the mass reconstruction for the median
particle $\rho^\pm$. The fitted mass value is  about 2.0\% lower
than the PDG value.

\end{multicols}
\ruleup
\begin{center}
\begin{minipage}[t]{0.4\textwidth}
\centering
\begin{overpic}[scale=0.38]{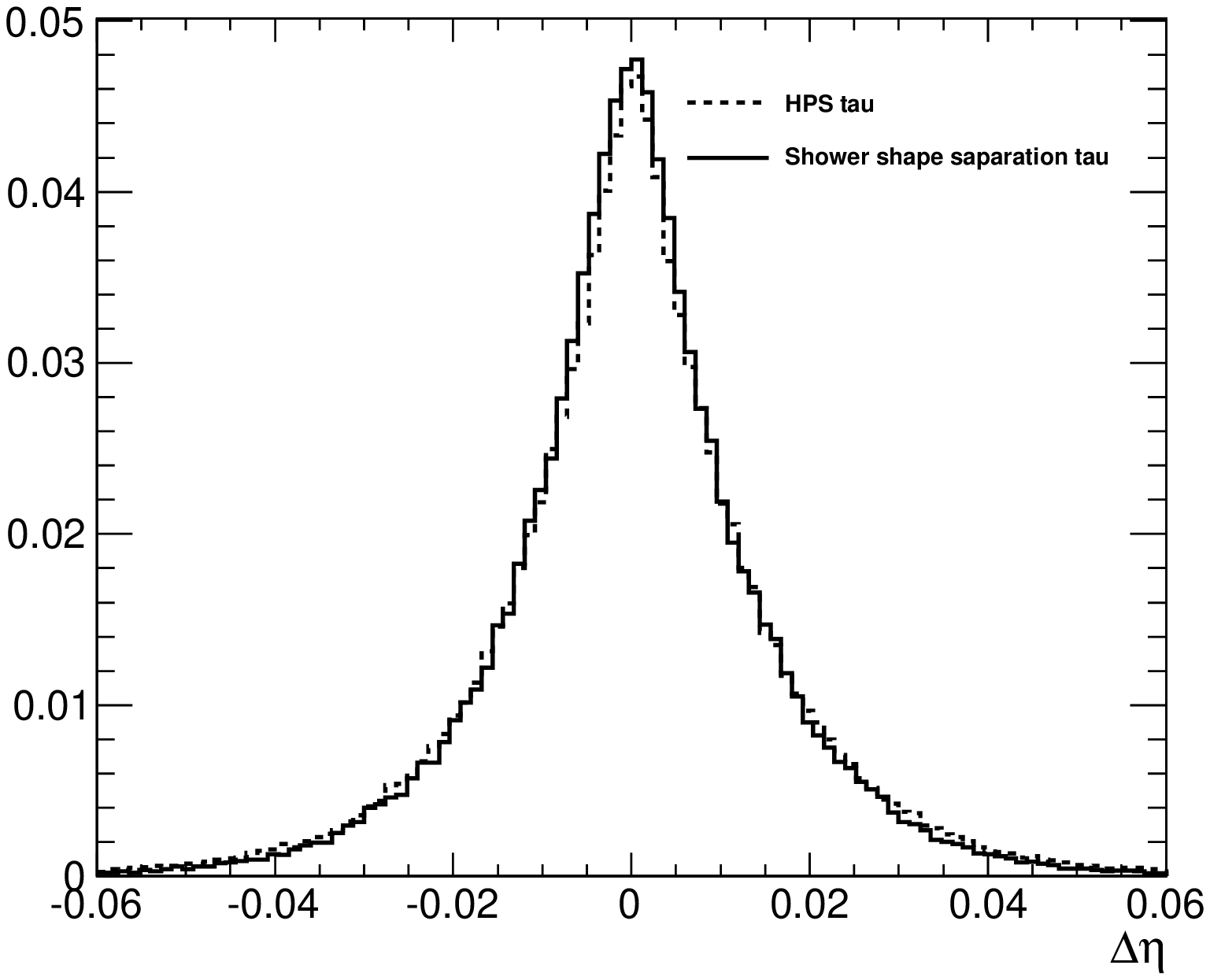}
\put(90,0){$(a)$}
\end{overpic}
\end{minipage}
\begin{minipage}[t]{0.4\textwidth}
\centering
\begin{overpic}[scale=0.38]{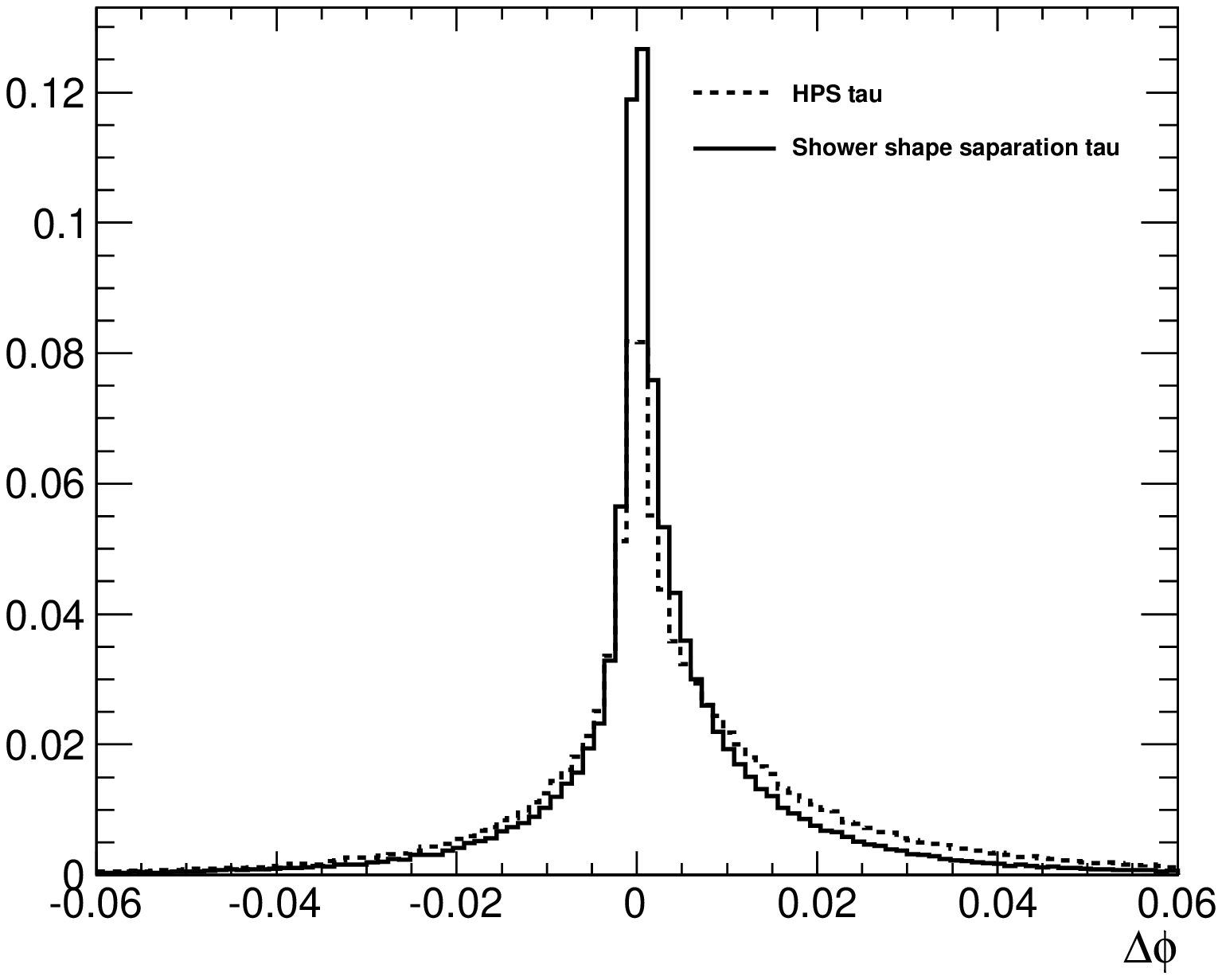}
\put(90,0){$(b)$}
\end{overpic}
\end{minipage}
\begin{minipage}[t]{0.4\textwidth}
\centering
\begin{overpic}[scale=0.38]{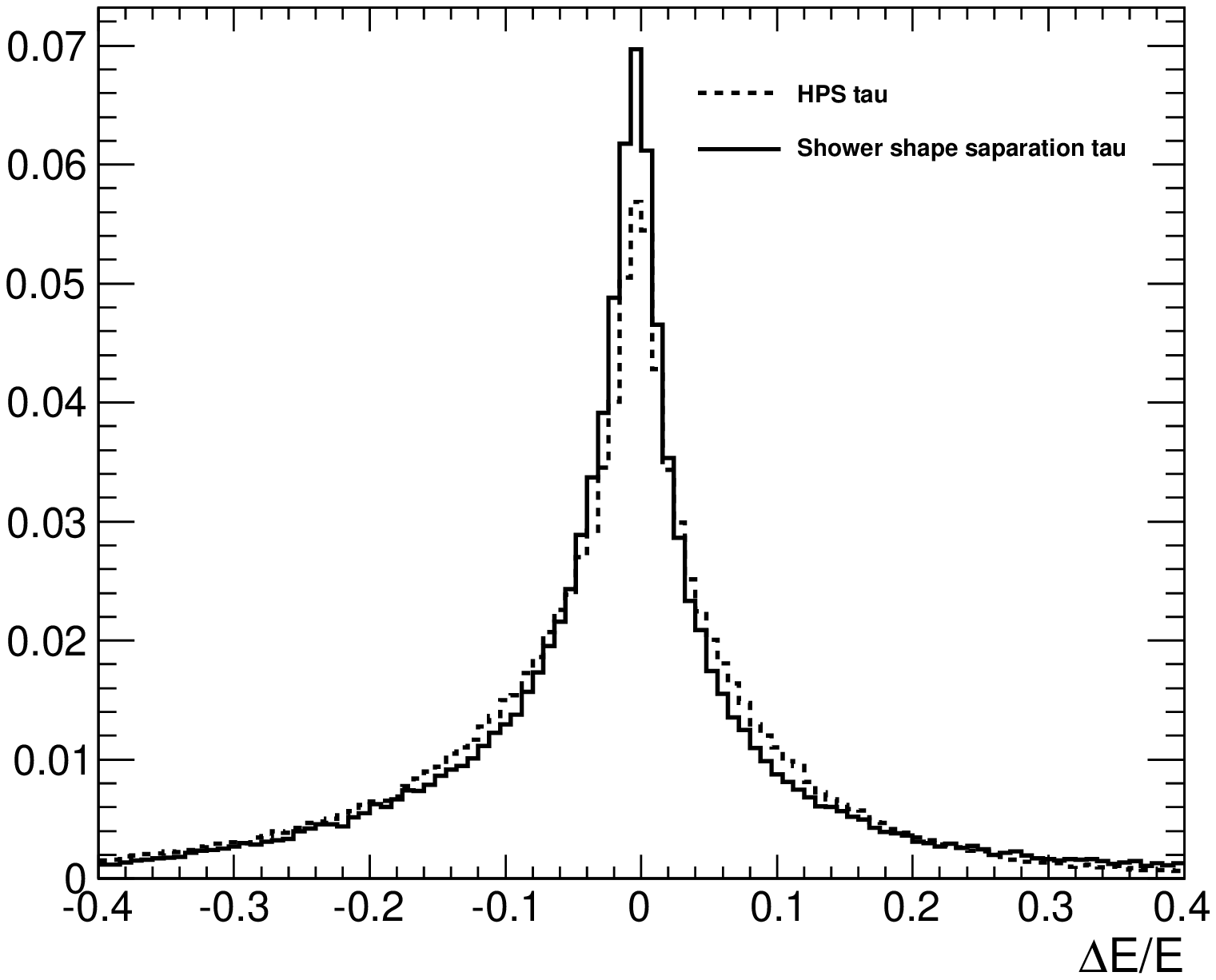}
\put(90,0){$(c)$}
\end{overpic}
\end{minipage}
\begin{minipage}[t]{0.4\textwidth}
\centering
\begin{overpic}[scale=0.38]{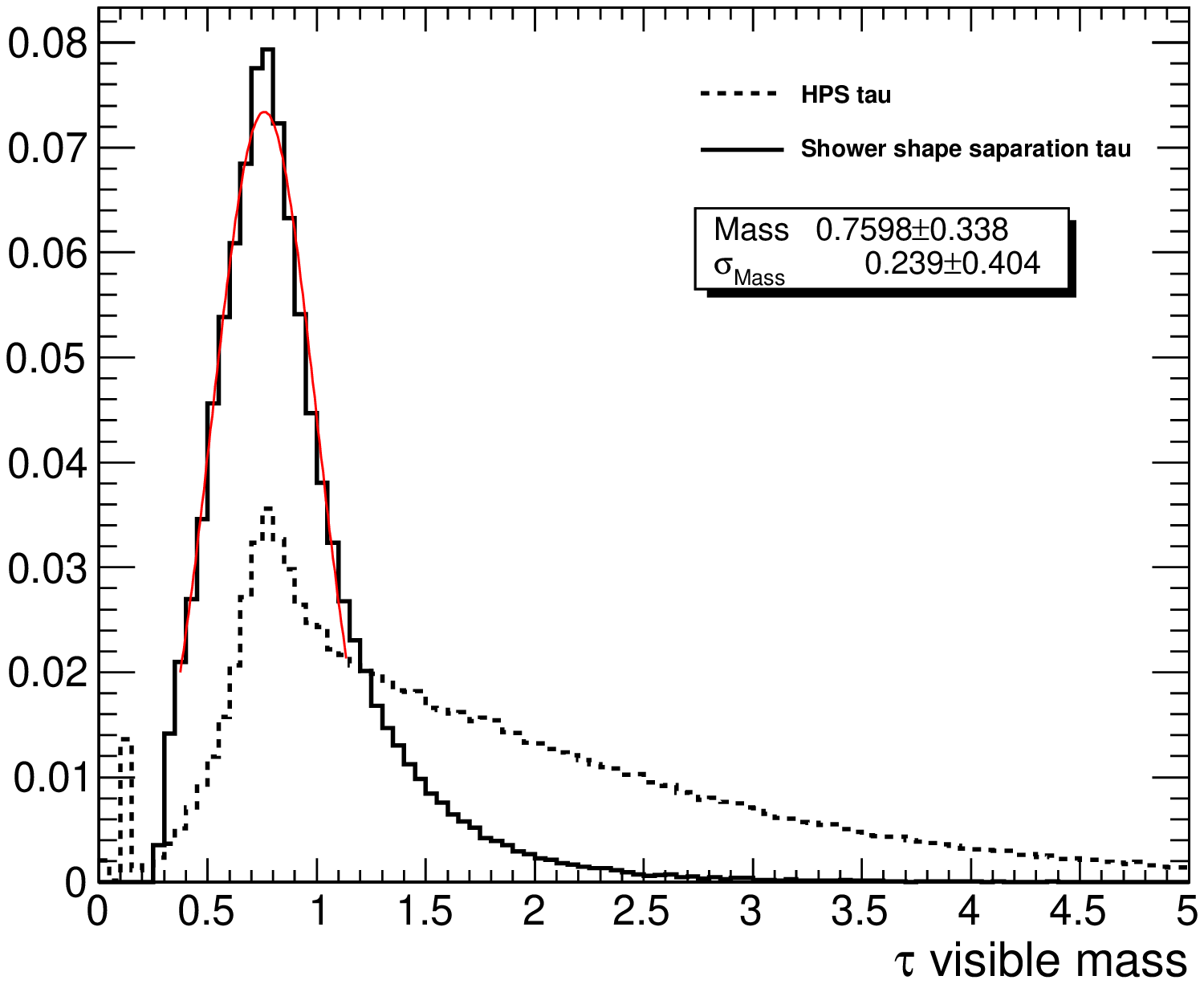}
\put(90,0){$(d)$}
\end{overpic}
\end{minipage}
\figcaption{The performance of $\rho^\pm$ reconstruction after the
application of the proposed method in this paper. (a) The
differences of position in $\eta$ and (b) $\phi$ directions,  and
(c) the difference on the energy $\frac{\Delta E}{E}$ between the
reconstructed $\rho$ with the separation method and the truth one.
(d) The visible mass of reconstructed
$\tau^\pm$($\rho^\pm$).}{\label{fig5}}
\end{center}
\ruledown

\begin{multicols}{2}

\section{Summary and outlook}
In this paper, the parametric formula of hadronic shower was used
for the substraction of the energy deposited by the charged hadron
from the overlapped shower from a charged hadron and a neutral
hadron in ECAL.  This is the first time to use the shower shape
method as one of the applications of the parametric formulae of
hadron shower. Taking the hadronic decay of tau,
$\tau^{\pm}\to\rho^{\pm}+\nu_{\tau^{\pm}}\to\pi^{\pm}+\pi^0+\nu_{\tau^{\pm}}$
for example,  the energy and position for the neutral pion can be
reconstructed satisfactorily after the separation with the proposed
technique in this paper. Finally improved results of the position,
energy and mass reconstructions of the median particle $\rho$ were
obtained, comparing with the present algorithm used in CMS at the
LHC.

In this paper, we take only the one prong hadronic decay of tau for
example to describe the method we proposed. In the near future, this
method will be studied to be used in the tau analysis with
multiplicity decay including several charged pions and neutral
pions. And this technique can be also used in the jet reconstruction
which also include the shower overlapping problem between charged
hadrons and neutral hadrons in ECAL. The reconstructed resolutions
both of position and energy of the jet and even the missing
transverse energy (MET) in experiment can be improved in the future,
which is very important in many physics analysis at the LHC. We also
noticed that the reconstructed positions in $\phi$ direction have
asymmetry distributions, due to the effect from the constructed
magnetic field with 3.8 Tesla along the $\eta$ direction. The effect
from the magnetic field and the approximate process can also induce
the asymmetry distribution of the reconstructed energy of the
particle $\rho$ in Fig.~\ref{fig5}(c). In the future analysis, the
correction from the magnetic field and the re-optimization of the
approximations should be considered.

\end{multicols}

\vspace{-1mm}
\centerline{\rule{80mm}{0.1pt}}
\vspace{2mm}

\begin{multicols}{2}

\end{multicols}

\clearpage

\end{CJK*}
\end{document}